# Latent Function Decomposition for Forecasting Li-ion Battery Cells Capacity: A Multi-Output Convolved Gaussian Process Approach

Abdallah A. Chehade[1] and Ala A. Hussein[2]

***Abstract*—A latent function decomposition method is proposed for forecasting the capacity of lithium-ion battery cells. The method uses the Multi-Output Gaussian Process, a generative machine learning framework for multi-task and transfer learning. The MCGP decomposes the available capacity trends from multiple battery cells into latent functions. The latent functions are then convolved over kernel smoothers to reconstruct and/or forecast capacity trends of the battery cells. Besides the high prediction accuracy the proposed method possesses, it provides uncertainty information for the predictions and captures nontrivial cross-correlations between capacity trends of different battery cells. These two merits make the proposed MCGP a very reliable and practical solution for applications that use battery cell packs. The MCGP is derived and compared to benchmark methods on an experimental lithium-ion battery cells data. The results show the effectiveness of the proposed method.**

## NOMENCLATURE

| | |
|---|---|
| *BMS* | Battery cell management system. |
| *EV* | Electric vehicle. |
| *CC* | Constant-current. |
| *CV* | Constant voltage. |
| *Li-ion* | Lithium-ion. |
| *MAE* | Mean-absolute error. |
| *MSE* | Mean-squared error. |
| *RUL* | Remaining useful life. |
| *SOC* | State-of-charge. |
| *SOH* | State-of-health. |
| *GP* | Gaussian process. |
| *MGP* | Multi-output Gaussian process. |
| *MCGP* | Multi-output convolved Gaussian process. |
| *IGP* | Independent Gaussian process. |
| *ANN* | Artificial neural network. |
| *RNN* | Recurrent neural network. |
| *BLUE* | Best linear unbiased estimator. |

[1] Department of Industrial and Manufacturing Systems Engineering at University of Michigan-Dearborn, Dearborn, Michigan, USA, achehade@umich.edu.

[2] Department of Electrical and Computer Engineering at University of Central Florida, Orlando, USA.

## I. INTRODUCTION

EVs still have many challenges that have not been fully addressed such as those related to battery performance prediction and aging/degradation modeling. Battery degradation is an extremely sophisticated process that involves highly nonlinear dynamics that are uneasy to predict ahead-of-time. As the battery ages, its capacity and power capability decrease due to the deterioration of its active materials. The capacity of a battery is a key performance parameter that must be tracked closely through the service-life of the battery [1], [2]. The capacity is defined as the maximum energy the battery can hold at a specific temperature. By definition, the capacity is directly correlated to the battery SOH and SOC [2]. Hence, a direct benefit of improving capacity predictions is improving the SOC and SOH estimation accuracy. Furthermore, accurate capacity forecasting allows for reliable RUL (in EVs, the battery usually reaches its service-life when its capacity drops by 20%). Accordingly, developing an accurate and reliable capacity estimation is also vital for maximizing the performance of EVs and developing robust BMS.

The research on capacity predictions for Li-ion batteries is rapidly growing and many methods are proposed. We classify those methods into two main categories: physics-based models and data-driven models. Physics-based models rely on a dynamic model that describes the physical/electrical behavior of the battery cell [3]–[5]. Data-driven models rely on machine learning and/or advanced mathematical/statistical models such as those proposed in [6]–[11]. In summary, most of the existing approaches are designed for (i) interpolation such as parametric models and Gaussian processes and (ii) short-term extrapolation such as neural networks. More details on commonly used machine learning approaches are discussed in Section II. However, interpolation and short-term extrapolation are not sufficient to develop a robust BMS. Therefore, there is still a need to investigate advanced non-parametric machine learning models that are reliable and robust for long term capacity forecasting.

To address this need, a Multi-output Convolved Gaussian Process (MCGP) [12]–[19] is proposed to forecast the capacity trend of battery cells that are relatively at an early degradation state. The MCGP decomposes the available capacity trends from the available battery cells into latent functions using the convolution process. The latent functions are then convolved with kernel smoothers to reconstruct and/or extrapolate the capacity trends of the available battery cells. The concept is similar to low-rank matrix decompositions (e.g., singular value decompositions) for matrix completion and recommender systems [20]–[25].

The contributions of the MCGP framework are multifold: (i) it simultaneously models the capacity of multiple battery cells, which enables multi-task learning and transferring knowledge between battery cells. (ii) It provides uncertainty information around the predictions, which serves as a robustness-metric. (iii) It is non-parametric and does not force a specific function shape on the capacity trends. (iv) It learns nontrivial cross-correlations between the battery cells.

The remainder of the paper is organized as follows: Section II summarizes the literature review. Section III discusses the proposed MCGP. Section IV validates the predictive performance of the proposed MCGP in comparison to benchmark models on experimental battery cells datasets [31]. Finally, Section V concludes the paper.

## II. Literature Review

Capacity trends are influenced by aging factors, manufacturing stochasticity, and various operational and environmental conditions (e.g., temperature and current). Under the assumption that the cells are operating at similar current/voltage profiles and the manufacturing process is highly controlled, the challenge is to capture the aging and environmental effects. The aging effect can be partially captured by modeling the correlation between the capacity and the charging-discharging cycle number. Therefore, the remaining challenge is capturing the dynamic environmental (most importantly temperature) effect on the capacity profile. Below, we discuss some commonly used machine learning approaches that can be leveraged for modeling capacity trends.

### A. *Artificial Neural Networks*

Artificial Neural Networks (ANNs) are commonly used to capture the nonlinear relation between a set of inputs and an output of interest [6], [26], [27]. Fig. 1 summarizes a framework for modeling capacity trends via ANN.

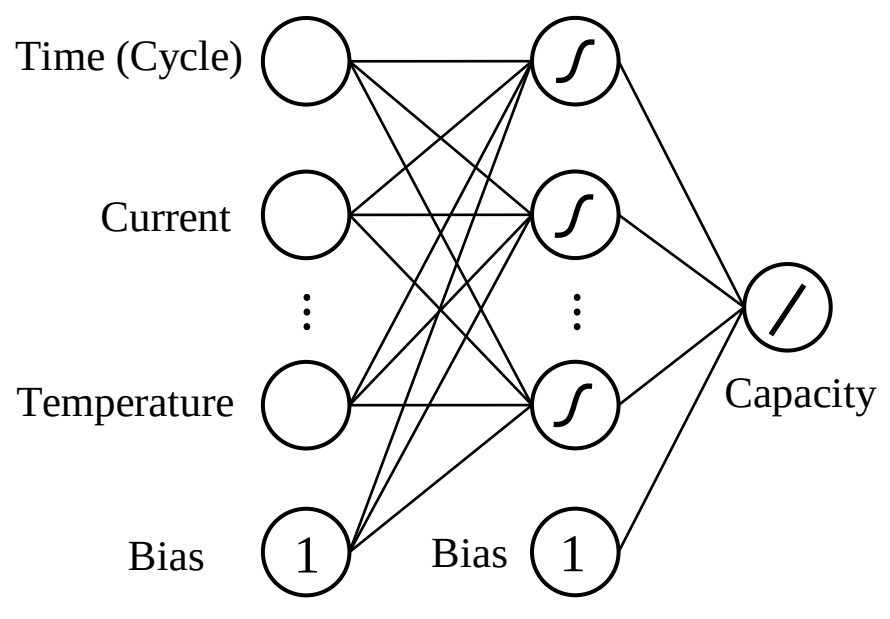

Fig. 1. An Artificial feedforward neural network topology for modeling the capacity of battery cells.

To extrapolate the capacity trend, ANNs require also extrapolating the input profiles, which is extremely challenging by itself. One option that avoids the extrapolation of the operational and environmental conditions is using historical environmental and operational profiles. The uncertainty of the input profiles tend to increase the uncertainty of the capacity predictions, which significantly accumulate for long term extrapolation. ANNs are trained to maximize the average performance and not the individualistic cell performance of each cell. Furthermore, ANNs ignore the cell-to-cell variations which also increase the uncertainty of the capacity predictions.

### B. *Recurrent Neural Networks*

Recurrent Neural Networks (RNNs) are also common to model time series trends [28], [29]. Fig. 2 summarizes one possible RNN framework for modeling capacity of battery cells. RNNs are expected to be more robust than ANNs for short term extrapolation with the support of the capacity feedback loop. Similar to ANNs, the RNNs require extrapolating the input profiles.

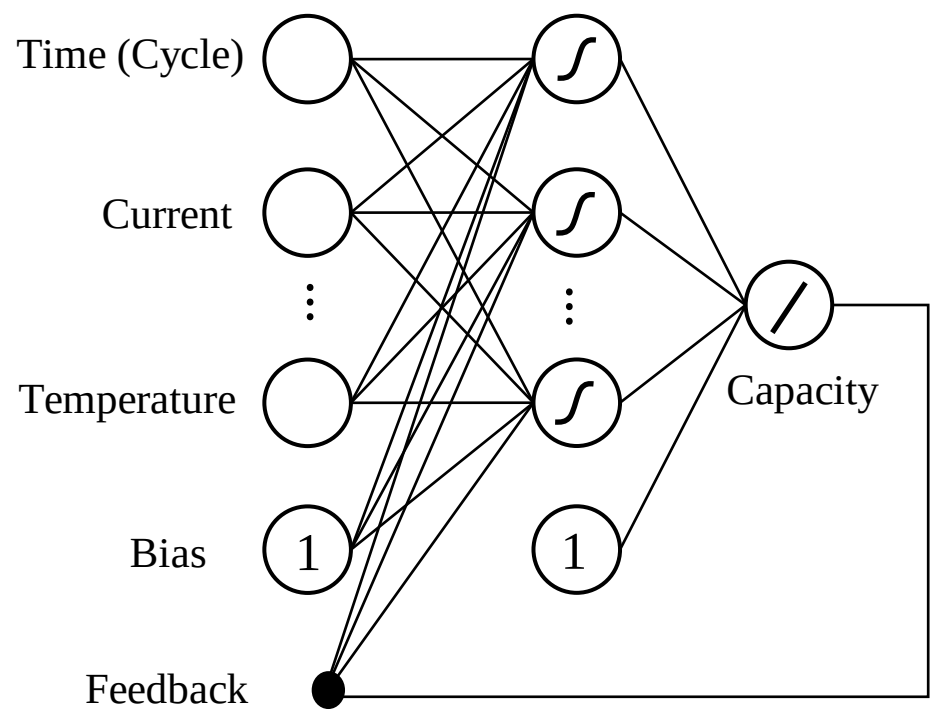

Fig. 2. A Recurrent neural network topology for modeling the capacity of battery cells.

While RNNs considers the previously measured/estimated capacity, it is still not robust for long term extrapolation due to the accumulated capacity prediction errors.

### C. *Independent Gaussian Processes*

One challenge for neural networks is stability and overfitting because of the large number of parameters (neural network weights) to be estimated. To avoid overfitting, the neural network must be either (i) simplified to decrease the number of estimated parameters or (ii) more data should be provided to robustly estimate the neural network parameters. However, when simplifying the neural network, its capability to accurately model the capacity deteriorates.

An alternative nonparametric approach to neural networks that aims to find nontrivial correlations between the input and the output is the Gaussian Process (GP) [8], [30] . A GP can be developed to model each capacity measurement from battery cell $i$ as a random variable (i.e., distribution of estimations), which provides a metric to quantify the uncertainty (also robustness) of the GP predictions. Furthermore, GPs focus on modeling the correlation of the capacity measurements at different cycles. Fig. 3 summarizes the framework for capacity predictions via independent GPs.

$$y_i(t) \sim \mathcal{GP}\left(b_i(\boldsymbol{x}_t), k(\boldsymbol{\theta}_i, \boldsymbol{x}_t, \boldsymbol{x}_{t'})\right) \quad (1)$$

where $y_i(t)$ is the capacity measurement from battery cell $i$ at charging-discharging cycle $t$, $b_i(.)$ is the basis function for the capacity of battery cell $i$, $k(\boldsymbol{\theta}_i, \boldsymbol{x}_t, \boldsymbol{x}_{t'}) = \text{cov}(y_i(t), y_i(t'))$ is the covariance function that correlates the pairwise capacity measurements for battery cell $i$ at cycles $t$ and $t'$, and $\boldsymbol{\theta}_i$ is set of the kernel hyper-parameters.

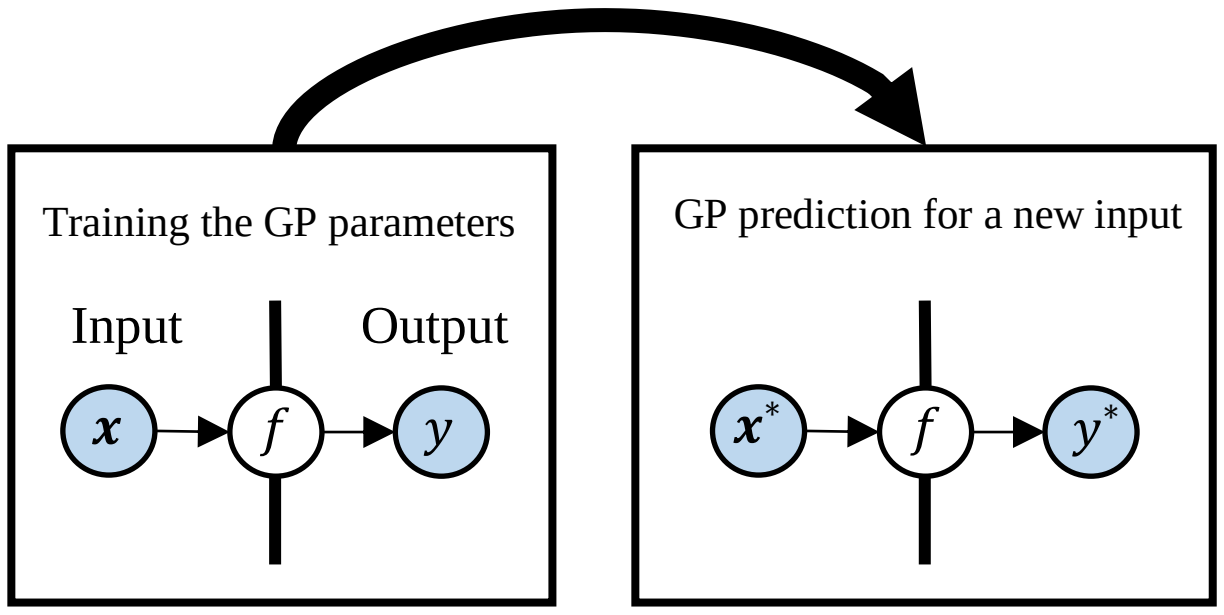

Fig. 3. The Independent GP model for capacity of battery cells.

The core assumption of GPs is that any subset of capacity measurements follow a normal distribution. Consequently, the joint distribution of the available capacity measurements $y_i(t_{i,1}), y_i(t_{i,2}), \dots, y_i(t_{i,n_i})$ follows the following multivariate normal distribution:

$$y_i(\boldsymbol{t}_i) = \begin{bmatrix} y_i(t_{i,1}) \\ y_i(t_{i,2}) \\ \vdots \\ y_i(t_{i,n_i}) \end{bmatrix} \sim \mathcal{N}\left( \begin{bmatrix} b_i(\boldsymbol{x}_{i,1}) \\ b_i(\boldsymbol{x}_{i,2}) \\ \vdots \\ b_i(\boldsymbol{x}_{i,n_i}) \end{bmatrix}, \boldsymbol{K}(\boldsymbol{\theta}_i, \boldsymbol{x}_{\boldsymbol{t}_i}, \boldsymbol{x}_{\boldsymbol{t}_i}) \right) \quad (2)$$

where

$$\boldsymbol{K}(\boldsymbol{\theta}_i, \boldsymbol{x}_{\boldsymbol{t}_i}, \boldsymbol{x}_{\boldsymbol{t}_i}) = \begin{bmatrix} k(\boldsymbol{\theta}_i, \boldsymbol{x}_{i,1}, \boldsymbol{x}_{i,1}) & k(\boldsymbol{\theta}_i, \boldsymbol{x}_{i,1}, \boldsymbol{x}_{i,2}) & \dots & k(\boldsymbol{\theta}_i, \boldsymbol{x}_{i,1}, \boldsymbol{x}_{i,n_i}) \\ k(\boldsymbol{\theta}_i, \boldsymbol{x}_{i,2}, \boldsymbol{x}_{i,1}) & \ddots & \dots & k(\boldsymbol{\theta}_i, \boldsymbol{x}_{i,2}, \boldsymbol{x}_{i,n_i}) \\ \vdots & \vdots & \ddots & \vdots \\ k(\boldsymbol{\theta}_i, \boldsymbol{x}_{i,n_i}, \boldsymbol{x}_{i,1}) & k(\boldsymbol{\theta}_i, \boldsymbol{x}_{i,n_i}, \boldsymbol{x}_{i,2}) & \dots & k(\boldsymbol{\theta}_i, \boldsymbol{x}_{i,n_i}, \boldsymbol{x}_{i,n_i}) \end{bmatrix}.$$

The kernel function depends on domain knowledge and the number of available observations. With enough available observations, the scaled Gaussian kernel captures the local trends in small time windows. For long term cyclic patterns, customized compound kernels can be developed that consists of seasonal kernels and/or seasonal mean functions. For the studied battery cell datasets, the scaled Gaussian kernel in (3) is shown to be effective. For the scaled Gaussian kernel, the prior covariance between the capacity measurements at cycles $t$ and $t'$ for battery cell $i$ can be written as

$$\text{cov}(y_i(t), y_i(t')) = k(\boldsymbol{x}_t, \boldsymbol{x}_{t'} | \theta_F, \theta_L, \theta_{x_1}, \theta_{x_2}, \dots, \theta_{x_N})$$
$$= \theta_F^2 \exp\left[ -\frac{1}{2}\left(\frac{t - t'}{\theta_L}\right)^2 - \frac{1}{2}\sum_{j=1}^{N}\left(\frac{x_{t,j} - x_{t',j}}{\theta_{x_j}}\right)^2 \right]. \quad (3)$$

Typically the basis function is defined to be $b_i(\boldsymbol{x}_t) = 0$ and the covariance kernel hyper-parameters $\boldsymbol{\theta}_i$ are determined to maximize the log-likelihood in eq. (4).

$$\log(p(\boldsymbol{y}_i(\boldsymbol{t}_i) | \boldsymbol{\theta}_i)) = -\frac{1}{2}\boldsymbol{y}_i(\boldsymbol{t}_i)^T \boldsymbol{K}(\boldsymbol{\theta}_i, \boldsymbol{x}_{\boldsymbol{t}_i}, \boldsymbol{x}_{\boldsymbol{t}_i})^{-1} \boldsymbol{y}_i(\boldsymbol{t}_i) - \frac{1}{2}\log\left(\det\left(\boldsymbol{K}(\boldsymbol{\theta}_i, \boldsymbol{x}_{\boldsymbol{t}_i}, \boldsymbol{x}_{\boldsymbol{t}_i})\right)\right) - \frac{n_i}{2}\log(2\pi) \quad (4)$$

A GP is fully defined with the maximum likelihood hyper-parameters and its basis function. Consider now an unobserved capacity measurement $y_i(t)$; it must be normally distributed according to the GP core assumption. Furthermore, the joint distribution of the observed measurements $\boldsymbol{y}_i(\boldsymbol{t}_i)$ and $y_i(t)$ must also follow a multivariate normal distribution according to the GP core assumption. Specifically, the joint distribution of the capacity measurements can be written as

$$\begin{bmatrix} \boldsymbol{y}_i(\boldsymbol{t}_i) \\ y_i(t) \end{bmatrix} \sim \left( \begin{bmatrix} \boldsymbol{0} \\ 0 \end{bmatrix}, \begin{bmatrix} \boldsymbol{K}(\boldsymbol{\theta}_i, \boldsymbol{x}_{\boldsymbol{t}_i}, \boldsymbol{x}_{\boldsymbol{t}_i}) & \boldsymbol{k}(\boldsymbol{\theta}_i, \boldsymbol{x}_t, \boldsymbol{x}_{\boldsymbol{t}_i}) \\ \boldsymbol{k}(\boldsymbol{\theta}_i, \boldsymbol{x}_{\boldsymbol{t}_i}, \boldsymbol{x}_t) & k(\boldsymbol{\theta}_i, \boldsymbol{x}_t, \boldsymbol{x}_t) \end{bmatrix} \right) \quad (5)$$

where $\boldsymbol{k}(\boldsymbol{\theta}_i, \boldsymbol{x}_t, \boldsymbol{x}_{\boldsymbol{t}_i}) = [k(\boldsymbol{\theta}_i, \boldsymbol{x}_t, \boldsymbol{x}_{i,1}) \quad \dots \quad k(\boldsymbol{\theta}_i, \boldsymbol{x}_t, \boldsymbol{x}_{i,n_i})]$.

Given the prior information $\boldsymbol{y}_i(\boldsymbol{t}_i), \boldsymbol{x}_t, \boldsymbol{x}_{\boldsymbol{t}_i}, \boldsymbol{\theta}_i$, the only unknown is $y_i(t)$ and its predictive posterior distribution is:

$$\hat{y}_i(t) | \boldsymbol{y}_i(\boldsymbol{t}_i), \boldsymbol{x}_t, \boldsymbol{x}_{\boldsymbol{t}_i}, \boldsymbol{\theta}_i \sim \mathcal{N}\left(\hat{\mu}_i(t), \hat{\sigma}_i^{\,2}(t)\right) \quad (6)$$

where

$$\hat{\mu}_i(t) = \boldsymbol{k}(\boldsymbol{\theta}_i, \boldsymbol{x}_t, \boldsymbol{x}_{\boldsymbol{t}_i}) \boldsymbol{K}(\boldsymbol{\theta}_i, \boldsymbol{x}_{\boldsymbol{t}_i}, \boldsymbol{x}_{\boldsymbol{t}_i})^{-1} y_i(\boldsymbol{t}_i) \quad (7)$$

and

$$\hat{\sigma}_i^{\,2}(t) = k(\boldsymbol{\theta}_i, \boldsymbol{x}_t, \boldsymbol{x}_t) - \boldsymbol{k}(\boldsymbol{\theta}_i, \boldsymbol{x}_t, \boldsymbol{x}_{\boldsymbol{t}_i}) \boldsymbol{K}(\boldsymbol{\theta}_i, \boldsymbol{t}_i, \boldsymbol{t}_i)^{-1} \boldsymbol{k}(\boldsymbol{\theta}_i, \boldsymbol{x}_{\boldsymbol{t}_i}, \boldsymbol{x}_t). \quad (8)$$

The concept is generic for other choices of kernels. For example, it is straight forward to add a noise term to the covariance function as the following:

$$\text{cov}(y_i(t), y_i(t')) = k(\boldsymbol{x}_t, \boldsymbol{x}_{t'} | \theta_\epsilon, \theta_F, \theta_L, \theta_{x_1}, \theta_{x_2}, \dots, \theta_{x_N})$$
$$= \theta_F^2 \exp\left[ -\frac{1}{2}\left(\frac{t - t'}{\theta_L}\right)^2 - \frac{1}{2}\sum_{j=1}^{N}\left(\frac{x_{t,j} - x_{t',j}}{\theta_{x_j}}\right)^2 \right] + \theta_\epsilon^2 \delta_{tt'} \quad (9)$$

where $\delta_{tt'}$ is the Kronecker delta function and it is equal to 1 if $t = t'$ and 0 otherwise.

The major limitation of IGPs for time-series modeling is extrapolation (i.e., forecasting). This is mainly because the basis model is usually considered to be 0 or a poor approximation of the true capacity trend, which is unknown.

Furthermore, the differences between the basis model and the true capacity trend accumulate for long term forecasting where there is no neighboring capacity data.

Therefore, IGPs are not suitable for capacity extrapolation. However, they provide a unique framework that can be extended to cross-correlate capacity trends of different battery cells, which supports extrapolating capacity trends for newly monitored battery cells.

### D. Multi-Output Gaussian Processes

Multi-Output Gaussian Processes (MGPs) are an extension to IGPs that consider the cross-correlations between multiple outputs [30]. Here, the multiple outputs are the capacity trends from the available battery cells. Therefore, an MGP for modeling capacity trends consider (i) correlations between different cycles of the same battery cell, (ii) cross-correlations between different cycles of different battery cells, and (iii) cross-correlations between similar cycles of different battery cells.

The challenge is to effectively model the cross-correlations between different battery cells. For this paper, we propose a novel Multi-Output Convolved Gaussian Process (MCGP) to decompose the available capacity trends from multiple Li-ion battery cells into multiple latent functions. The latent functions will then be convolved with optimized kernel smoothers to extrapolate the capacity trend of the battery cell of interest. The proposed MCGP allows for transfer learning through the learned latent functions and also serve as a multi-task learner that simultaneously model the capacity trends of multiple battery cells. Fig. 4 summarizes the MCGP framework for capacity predictions via MCGPs.

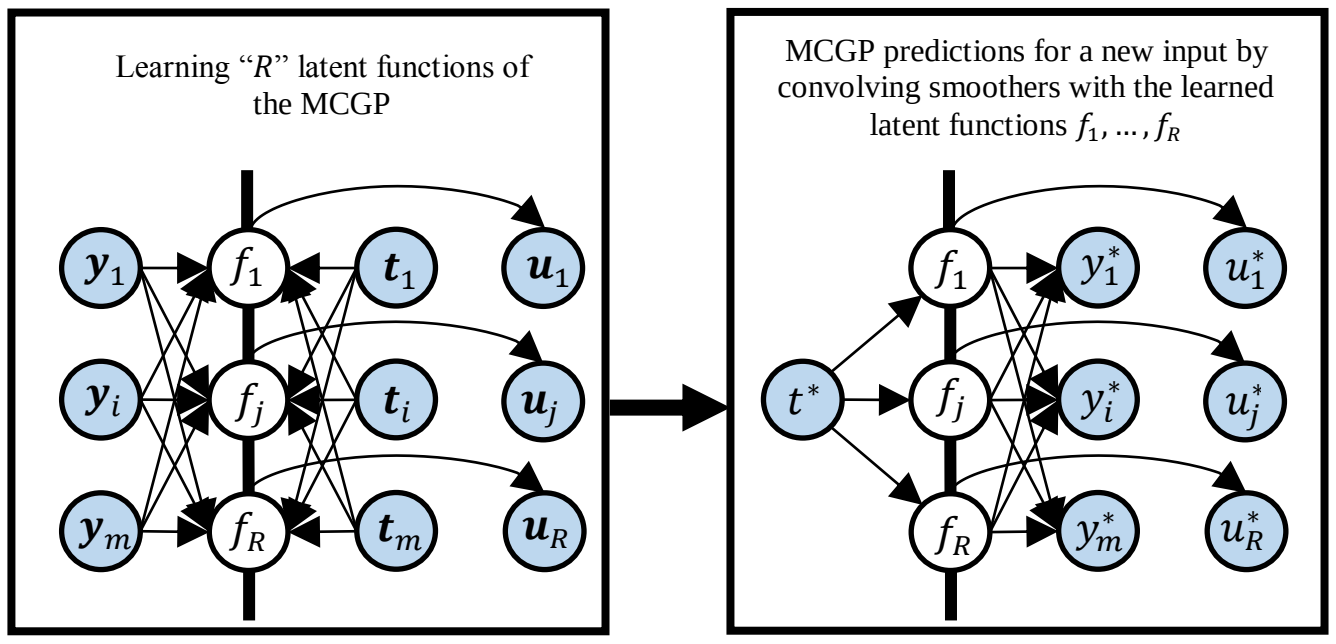

Fig. 4. The proposed MCGP model for predicting the capacity at a new cycle $t^*$ of battery cells $1, \dots, m$ including the cell of interest $i$.

## III. Multi-Output Convovled Gaussian Process for Modeling Capacity Trends of Baterry Cells

### A. Conceptual Idea

For the MCGP framework, the capacity trends of the available battery cells are assumed to share some function similarities. In other words, it is assumed that the capacity from the available battery cells can be decomposed into latent functions. The latent functions will then be leveraged to generate capacity trends for new inputs, which allows to extrapolate the capacity trend of newly monitored battery cells. The concept is similar to matrix decomposition for matrix completion. First, the incomplete matrix is decomposed into two low-rank matrices and then the low-rank matrices are leveraged to complete the full matrix.

Recall that considering current, temperature, and other environmental and operational conditions require forecasting them for accurate capacity extrapolation. To eliminate additional sources of uncertainty from such extrapolations, we consider the cycle number as the only input. The assumption here is that the operational and environmental effects are already intrinsically embedded in the capacity profile of a battery cell. Therefore, the cross-correlations between the capacity trends of different battery cells are assumed to also capture the cross-correlations between the operational and environmental conditions of the cells.

### B. Methodology Development

Given the capacity measurements from the $m$ available battery cells by $\left\{\left(t_{i,k}, y_i\left(t_{i,k}\right)\right); k = 1,2, \dots, n_i \;\&\; i = 1,2., \dots m\right\}$. Here, $n_i$ is the number of observed cycles for cell $i$, $t_{i,k}$ is the cycle of the $k^{th}$ capacity measurement for cell $i$, and $y_i\left(t_{i,k}\right)$ is the capacity measurement at time $t_{i,k}$ for cell $i$.

Assuming that the available battery cells can be described by $R$ latent functions. The capacity at time $t$ for battery cell $i$ is expressed as a convolution between the latent functions $u_r(.)$ and smoothing kernels $k_{ir}(.)$ corrupted with a white noise $e(.)$.

$$y_i(t) = \sum_{r=1}^{R} \int_{-\infty}^{+\infty} k_{ir}(t-\Delta) u_r(\Delta) \mathrm{d}\Delta + e(t) \tag{10}$$

Each of the latent functions is assumed to follow an IGP

$$u_r(\Delta, \Delta^*) \sim \mathcal{GP}\left(0, k_r(\Delta, \Delta^*)\right) \tag{11}$$

and the noise is assumed to follow a normal distribution

$$e(t) \sim \mathcal{N}(0, \theta_\epsilon^2). \tag{12}$$

Under the independence assumption between the noise and the latent functions, the cross-covariance between the capacity of battery cell $i$ and battery cell $j$ can be written as

$$\mathrm{cov}\left[y_i(t), y_j(t^*)\right] = \theta_\epsilon^2 \delta_{tt^*} + \sum_{r=1}^{R} \sum_{p=1}^{R} \int_{-\infty}^{+\infty} k_{ir}(t-\Delta) \int_{-\infty}^{+\infty} k_{jp}(t^*-\Delta^*) \mathrm{cov}\left[u_r(\Delta), u_p(\Delta^*)\right] \mathrm{d}\Delta \mathrm{d}\Delta^* \tag{13}$$

where $\delta$ is the Kronecker delta function.

Furthermore and similar to the principal component analysis scheme, the latent functions are assumed to be independent. Under this consideration, the covariance of the latent functions can then be written as

$$\operatorname{cov}\left[u_r(\Delta), u_p(\Delta^*)\right] = k_r(\Delta, \Delta^*)\delta_{rp} \tag{14}$$

and the cross-covariance function in eq. (13) simplifies to

$$\operatorname{cov}\left[y_i(t), y_j(t^*)\right] = \theta_\epsilon^2 \delta_{tt^*} + \sum_{r=1}^{R} \int_{-\infty}^{+\infty} k_{ir}(t-\Delta) \int_{-\infty}^{+\infty} k_{jr}(t^*-\Delta^*) k_r(\Delta, \Delta^*) \mathrm{d}\Delta \mathrm{d}\Delta^*. \tag{15}$$

*C. Predictive Posterior Distribution*

Based on the MCGP, the prior distribution of the observed capacity values for the different available battery cells is

$$p(\boldsymbol{Y}|\boldsymbol{X}, \boldsymbol{\theta}) = \mathcal{N}\left(\boldsymbol{0}, \boldsymbol{K}_{\boldsymbol{X},\boldsymbol{X}}\right) \tag{16}$$

where $\boldsymbol{Y} = [\boldsymbol{y}_1(\boldsymbol{t}_1)^T, \dots, \boldsymbol{y}_N(\boldsymbol{t}_N)^T]^T$ is a $T$x1 vector of capacity measurements, $T = \sum_{i=1}^{N} n_{t_i}$, $\boldsymbol{y}_i(\boldsymbol{t}_i)$ is the available capacity data for battery cell $i$, $\boldsymbol{X} = [\boldsymbol{t}_1^T, \dots, \boldsymbol{t}_N^T]^T$ is a $T$x1 vector for the observed cycles, $\boldsymbol{\theta}$ is the set of hyper-parameters of the covariance function in eq. (15), and $\boldsymbol{K}_{\boldsymbol{X},\boldsymbol{X}}$ is a $T$x$T$ of the available measurements $\boldsymbol{Y}$ and it is calculated using eq. (15).

The estimated kernel parameter set $\boldsymbol{\theta}$ is the solution that maximizes the likelihood function of the prior distribution in eq. (4). More details on parameter estimation will be provided in Section III.D.

For capacity estimations, the predictive posterior distribution is considered in eq. (17).

$$p\left(\boldsymbol{c}_j(\boldsymbol{t}^*)|\boldsymbol{Y}, \boldsymbol{X}, \boldsymbol{\theta}\right) = \mathcal{N}\left(\boldsymbol{\mu}_j^*, \boldsymbol{\Sigma}_j^*\right) \tag{17}$$

where

$$\boldsymbol{\mu}_j^* = \boldsymbol{K}_{\boldsymbol{t}^*,\boldsymbol{X}}^{(j)} \boldsymbol{K}_{\boldsymbol{X},\boldsymbol{X}}{}^{-1} \boldsymbol{Y}$$

and

$$\boldsymbol{\Sigma}_j^* = \boldsymbol{K}_{\boldsymbol{t}^*,\boldsymbol{t}^*}^{(j)} - \boldsymbol{K}_{\boldsymbol{t}^*,\boldsymbol{X}}^{(j)} \boldsymbol{K}_{\boldsymbol{X},\boldsymbol{X}}{}^{-1} \boldsymbol{K}_{\boldsymbol{X},\boldsymbol{t}^*}^{(j)}.$$

Here, $\boldsymbol{K}_{\boldsymbol{t}^*,\boldsymbol{t}^*}^{(j)}$ and $\boldsymbol{K}_{\boldsymbol{t}^*,\boldsymbol{X}}^{(j)}$ are calculated by the covariance function in eq. (15).

The predictive distribution is the Best Unbiased Linear Estimate (BLUE). Furthermore, the predictive distribution is conditional on known information: (i) historical observed capacity data from the available battery cells, (ii) the time vector for forecasting, and (iii) the battery cell of interest. Note that the predictive distribution can be also leveraged for interpolation, smoothing, and imputing missing data. The MCGP framework is summarized in Fig. 4.

*D. Maximum Likelihood Estimator*

The predictive posterior distribution is conditional on the hyper-parameters of the covariance function in eq. (15). To learn the hyper-parameters, we aim to minimize the deviance (i.e., maximize the likelihood) in eq. (18).

$$\begin{aligned}
\mathcal{D}(\boldsymbol{\theta}|\boldsymbol{Y}, \boldsymbol{X}) &= -2\log\left(p(\boldsymbol{Y}|\boldsymbol{X}, \boldsymbol{\theta})\right) \\
&\propto \sum_{i=1}^{m} \boldsymbol{y}_i(\boldsymbol{t}_i)^T \boldsymbol{K}(\boldsymbol{\theta}_i, \boldsymbol{t}_i, \boldsymbol{t}_i)^{-1} \boldsymbol{y}_i(\boldsymbol{t}_i) \\
&\qquad + \log\left(\det\left(\boldsymbol{K}(\boldsymbol{\theta}_i, \boldsymbol{t}_i, \boldsymbol{t}_i)\right)\right) \\
&\propto \boldsymbol{Y}^T \boldsymbol{K}(\boldsymbol{\theta}, \boldsymbol{X}, \boldsymbol{X})^{-1} \boldsymbol{Y} + \log\left(\det\left(\boldsymbol{K}(\boldsymbol{\theta}, \boldsymbol{X}, \boldsymbol{X})\right)\right)
\end{aligned} \tag{18}$$

Most optimization algorithms including quasi-newton, trust-region, or gradient descent algorithms can be used to find the hyper-parameters that minimize the deviance (i.e., maximize the likelihood). Such optimization methods typically require estimating the gradient of the deviance with respect to the hyper-parameters. However, the determinate and the inverse of the kernel matrix make it challenging to calculate the gradient. Next, we will use the chain rule to find the gradient of the deviance with respect to the hyper-parameters.

First, the gradient can be written as

$$\frac{d\mathcal{D}(\boldsymbol{\theta}|\boldsymbol{Y}, \boldsymbol{X})}{d\boldsymbol{\theta}} = \underbrace{\frac{d\log\left(\det\left(\boldsymbol{K}(\boldsymbol{\theta}, \boldsymbol{X}, \boldsymbol{X})\right)\right)}{d\boldsymbol{\theta}}}_{(a)} - \boldsymbol{Y}^T \underbrace{\frac{\boldsymbol{dK}(\boldsymbol{\theta}, \boldsymbol{X}, \boldsymbol{X})^{-1}}{d\boldsymbol{\theta}}}_{(\boldsymbol{b})} \boldsymbol{Y} \tag{19}$$

where (a) can be expressed as

$$\frac{d\log\left(\det\left(\boldsymbol{K}(\boldsymbol{\theta}, \boldsymbol{X}, \boldsymbol{X})\right)\right)}{d\boldsymbol{\theta}} = \operatorname{trace}\left\{\boldsymbol{K}(\boldsymbol{\theta}, \boldsymbol{X}, \boldsymbol{X})^{-1} \frac{\boldsymbol{dK}(\boldsymbol{\theta}, \boldsymbol{X}, \boldsymbol{X})}{d\boldsymbol{\theta}}\right\}$$

and (b) can be expressed as

$$\frac{\boldsymbol{dK}(\boldsymbol{\theta}, \boldsymbol{X}, \boldsymbol{X})^{-1}}{d\boldsymbol{\theta}} = \boldsymbol{K}(\boldsymbol{\theta}, \boldsymbol{X}, \boldsymbol{X})^{-1} \frac{\boldsymbol{dK}(\boldsymbol{\theta}, \boldsymbol{X}, \boldsymbol{X})}{d\boldsymbol{\theta}} \boldsymbol{K}(\boldsymbol{\theta}, \boldsymbol{X}, \boldsymbol{X})^{-1}.$$

Accordingly, the gradient of the deviance with respect to the hyper-parameters can be also written as

$$\begin{aligned}
\frac{d\mathcal{D}(\boldsymbol{\theta}|\boldsymbol{Y}, \boldsymbol{X})}{d\boldsymbol{\theta}} &= \operatorname{trace}\left\{\boldsymbol{K}(\boldsymbol{\theta}, \boldsymbol{X}, \boldsymbol{X})^{-1} \frac{\boldsymbol{dK}(\boldsymbol{\theta}, \boldsymbol{X}, \boldsymbol{X})}{d\boldsymbol{\theta}}\right\} \\
&\quad - \frac{\left\{\boldsymbol{Y}^T \boldsymbol{K}(\boldsymbol{\theta}, \boldsymbol{X}, \boldsymbol{X})^{-1} \frac{\boldsymbol{dK}(\boldsymbol{\theta}, \boldsymbol{X}, \boldsymbol{X})}{d\boldsymbol{\theta}} \boldsymbol{K}(\boldsymbol{\theta}, \boldsymbol{X}, \boldsymbol{X})^{-1} \boldsymbol{Y}\right\}}{\boldsymbol{Y}^T \boldsymbol{K}(\boldsymbol{\theta}, \boldsymbol{X}, \boldsymbol{X})^{-1} \boldsymbol{Y}}.
\end{aligned} \tag{20}$$

With the use of the chain rule, the gradient of the deviance with respect to the hyper-parameters is now only dependent on the gradient of the kernel matrix with respect to the hyper-parameters, $\boldsymbol{dK}(\boldsymbol{\theta}, \boldsymbol{X}, \boldsymbol{X})/d\boldsymbol{\theta}$, as shown in eq. (20). $\boldsymbol{dK}(\boldsymbol{\theta}, \boldsymbol{X}, \boldsymbol{X})/d\boldsymbol{\theta}$ can be analytically or numerically calculated. Note that built-in numerical and automatic differentiation algorithms are available in most optimization toolboxes.

### E. Gaussian Kernels and Smoothers

The only remaining piece is to identify the kernel smoother in eq. (10) and the kernel covariance function in eq. (11) to calculate the gram kernel matrix using eq. (15). Without loss of generality, we follow the literature [16] and chose the following exponential smoother that relates the capacity of cell $i$ and latent function $r$

$$k_{ir}(t-\Delta) = \frac{\theta_{ir}^{(l)}}{\sqrt{2\pi\theta_{ir}^2}} \exp\left\{-\frac{1}{2}\left(\frac{t-\Delta}{\theta_{ir}}\right)^2\right\} \quad (21)$$

and the following Gaussian kernel for the IGP that describes latent function $r$

$$k_r(\Delta,\Delta^*) = \frac{1}{\sqrt{2\pi\theta_r^2}} \exp\left\{-\frac{1}{2}\left(\frac{\Delta-\Delta^*}{\theta_r}\right)^2\right\}. \quad (22)$$

The above-mentioned choices of kernels result in a closed form for the double integral in eq. (15) using the Gaussian distribution identity:

$$\int_{-\infty}^{+\infty} \frac{1}{\sqrt{2\pi\sigma^2}} \exp\left\{-\frac{1}{2}\left(\frac{\mu-\mu^*}{\sigma}\right)^2\right\} d\mu^* = 1. \quad (23)$$

## IV. Experimental Validation: NASA Dataset

### A. Dataset Description

The proposed MCGP is validated and compared to four other benchmark models using the battery datasets provided by [31]. The data comprise the voltage, the current, the temperature and the SOC of three battery cells (B0005, B0006, B0007) through repeated cycling tests starting with a new cell cycled until a 30% capacity-loss is recorded. Each charge-discharge cycle consists of a 1500-mA CC charging to 4.2-V followed by a CV charging at 4.2-V until the current drops to 20-mA, and 2000-mA CC discharging until the cell's voltage reaches its cutoff value (2.7-V, 2.5-V and 2.2-V for cells B0005, B0006 and B0007, respectively).

### B. MCGP Performance

For this case study, two latent functions were chosen to describe the shared information between the three battery cells. The latent functions are then used to regenerate, filter, and forecast the capacity trends of the three cells. More details on the results are shown in Fig. 5 and Table 1. Here, the proposed MCGP serve as a multi-task learner that simultaneously models the capacity trends of the three battery cells. It also supports transfer learning of trends between the cells. For training the MCGP, the data is uniformly downsampled for training (one of three consecutive capacity measurements are used for training). However, we did not downsample for capacity prediction (i.e., capacity forecasting, testing, and validating).

Fig. 5 (a) summarizes the raw capacity data of the first 100 cycles for cell B0005, the capacity data of the first 168 cycles for cells B0006 and B0007 in empty blue circles. The figure also shows the forecasted capacity for the remaining 68 cycles of cell B0005 in solid red and the true hidden capacity data for the remaining 68 cycles of cell B0005 in filled green circles. For this scenario, the training data is the capacity data for the first 168 cycles of cell B0006 and B0007 and capacity data for the first 100 cycles of cell B0005.

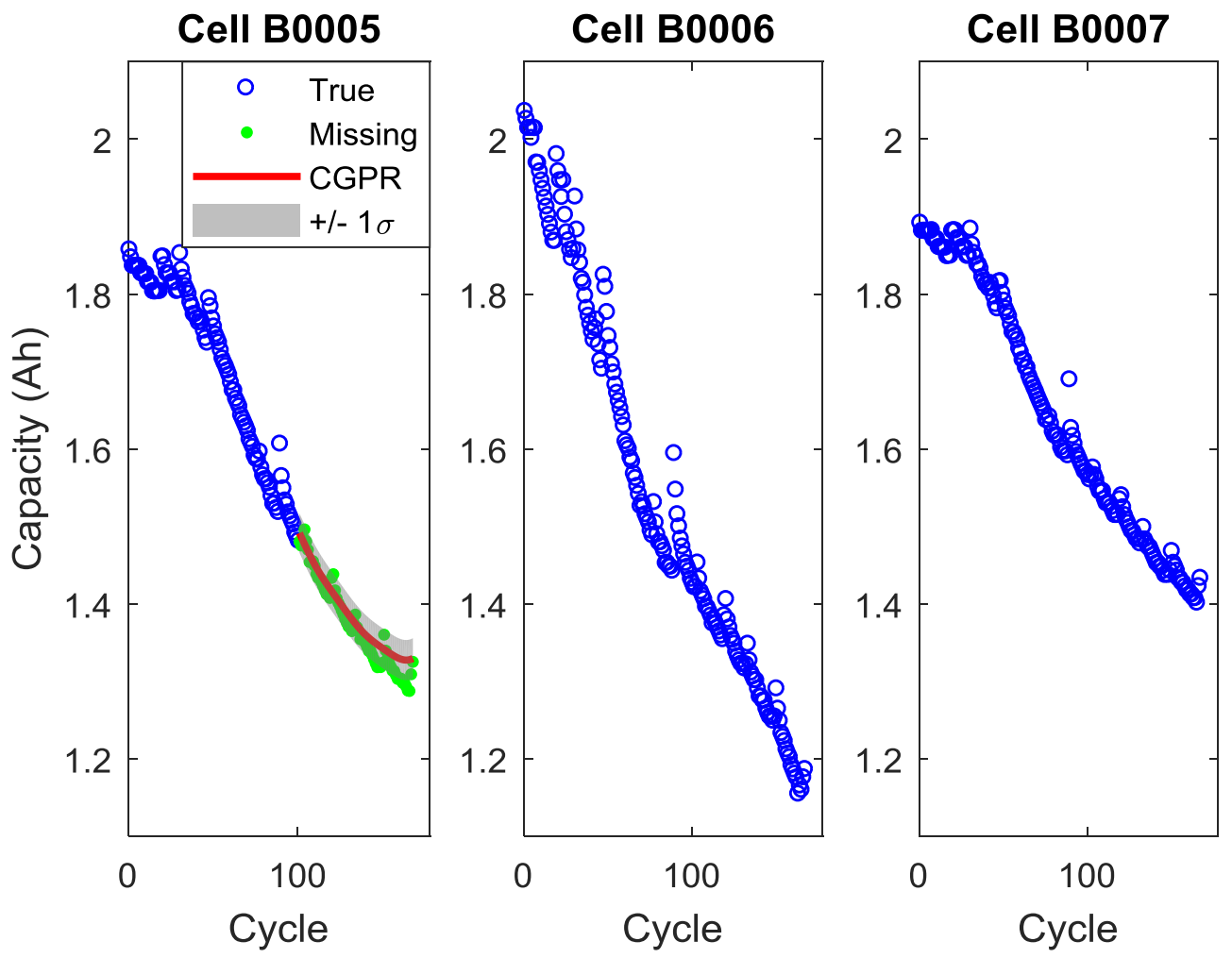

Fig. 5 (a) MCGP predictions for the raw capacity measurements of cell B0005.

Fig. 5 (a) shows that the MCGP successfully predicted the capacity for the remaining 68 cycles of cell B0005. The main reason behind this observation is the high cross-correlation between cell B0005 and the remaining cells. The figure also shows small variations around the predictions, which provides further evidence on the efficacy of the proposed MCGP.

Similarly, Fig. 5 (b) summarizes the forecasted capacities of CGPR of the intentionally hidden 68 cycles for cell B0007. Here, the training data is the raw capacity data for the first 168 cycles of cell B0005 and B0007 and the raw capacity data for the first 100 cycles of cell B0006.

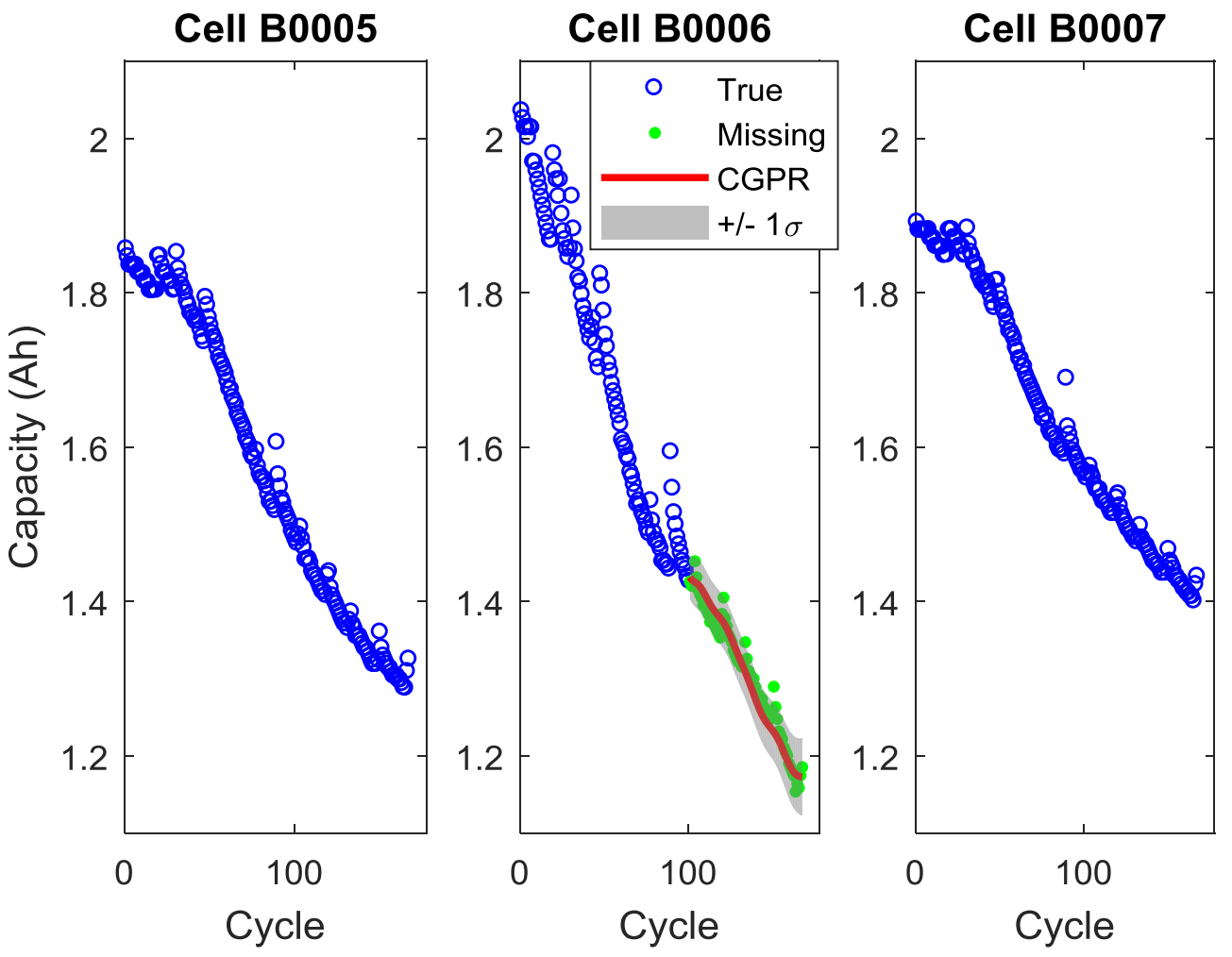

Fig. 5 (b) MCGP predictions for the raw capacity measurements of cell B0006.

It can be seen from Fig. 5 (b) that capacity profile for cell B0006 does not follow a similar trend as the other two cells. However, the MCGP still captured the nontrivial cross-correlations between cell B0006 and the other two cells and showed a similar effective performance as Fig. 5 (a). Capturing the nontrivial cross-correlations across cells is yet one major practical advantage of the MCGP.

Similarly, Fig. 5 (c) summarizes the forecasted capacities of CGPR of the intentionally hidden 68 cycles for cell B0007. Here, the training data is the raw capacity data for the first 168 cycles of cell B0005 and B0006 and the raw capacity data for the first 100 cycles of cell B0007. Fig. 5 (c) also shows an effective performance for the MCGP; however, the performance for cell B0007 slightly deteriorated compared to cell B0006 shown in Fig. 5 (b) and cell B0005 shown in Fig. 5 (a). This slight downgrade in the performance is mostly due to the discontinuous peak in capacity at cycle 89. Such discontinuity points are a major challenge for most machine learning models; however, the MCGP did partially overcome this challenge and captured the curvature in the trend.

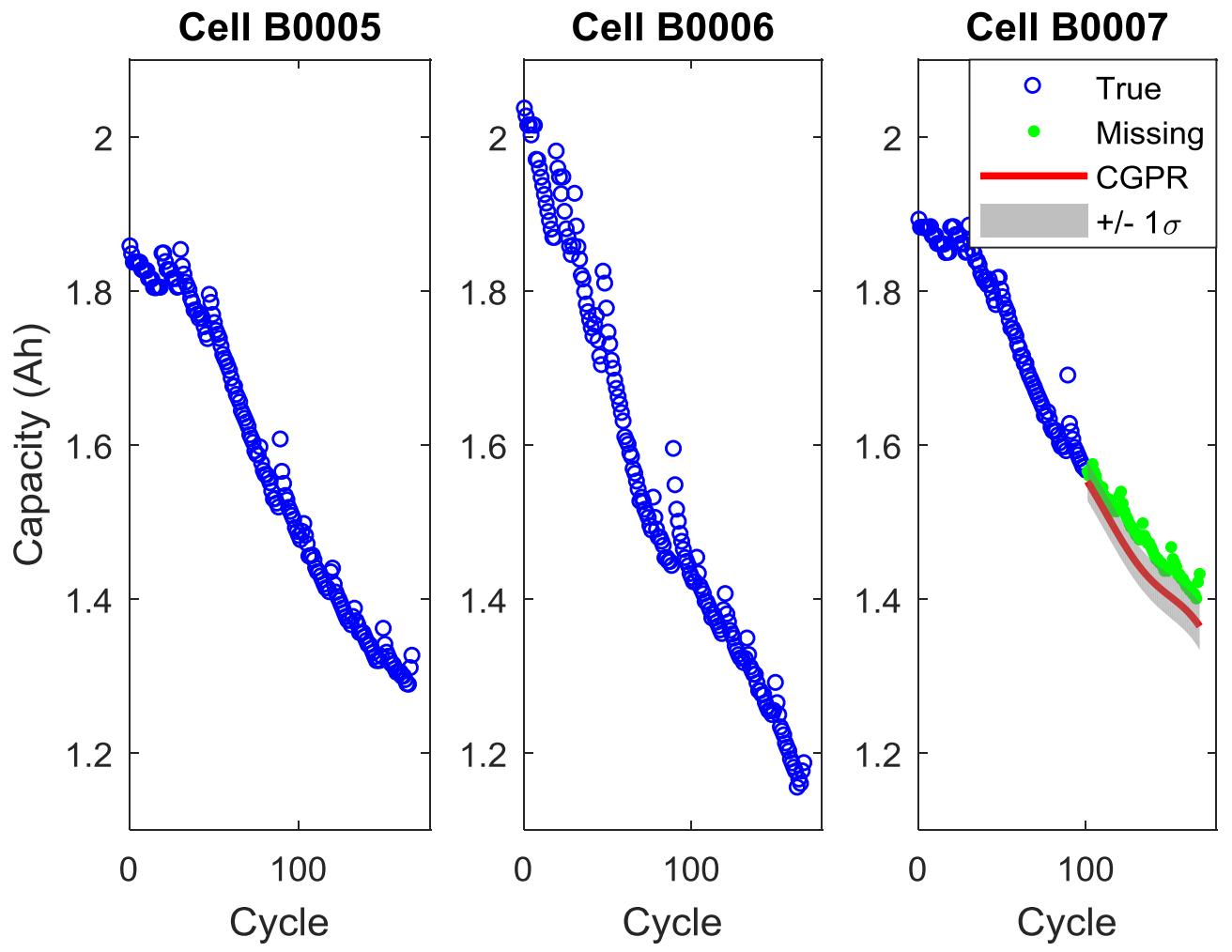

Fig. 5 (c) MCGP predictions for the raw capacity measurements of cell B0007.

Table 1 summarizes the kernel parameters in eq. (12), the training log-likelihood in eq. (4), and the training deviance in eq. (18) for the scenarios shown in Fig. 5. To follow-up with the subscripts of the parameters, please refer back to eq. (21) and eq. (22). Here, cells B0005, B0006, and B0007 are denoted by 1, 2, and 3, respectively. The values of estimated parameters summarizes the influence and connection between the capacity of the battery cells and the latent functions.

The table shows similar training log-likelihoods, and deviances of the three scenarios in Fig. (5). This is mainly because there is a large overlap between training capacity data of the three scenarios. Furthermore, this observation indicates that the MCGP training performance is similar for all three scenarios. However, this does not provide an indication about the prediction/forecasting performance of the MCGP and that will be investigated in the sub-section IV.C.

Table 1. Kernel parameters, log-likelihood and deviance for the scenarios in Fig. 5.

| **Scenario** | **Latent function** | **Fig. 5 (a)** | **Fig. 5 (b)** | **Fig. 5 (c)** |
|---|---|---|---|---|
| $\theta_{11}^{(l)}$ | 1 | 6.286x10$^{1}$ | 9.202x10$^{1}$ | 7.724x10$^{1}$ |
| $\theta_{21}^{(l)}$ | 1 | 3.246x10$^{1}$ | 7.256x10$^{1}$ | 6.746x10$^{1}$ |
| $\theta_{31}^{(l)}$ | 1 | 7.209x10$^{1}$ | 9.906x10$^{1}$ | 7.608x10$^{1}$ |
| $\theta_{11}$ | 1 | 5.240x10$^{0}$ | 4.134x10$^{0}$ | 4.851x10$^{0}$ |
| $\theta_{21}$ | 1 | 5.196x10$^{0}$ | 4.092x10$^{0}$ | 4.826x10$^{0}$ |
| $\theta_{31}$ | 1 | 5.355x10$^{0}$ | 4.240x10$^{0}$ | 4.927x10$^{0}$ |
| $\theta_{1}$ | 1 | 1.590x10$^{2}$ | 1.104x10$^{1}$ | 8.884x10$^{1}$ |
| $\theta_{12}^{(l)}$ | 2 | 1.240x10$^{1}$ | 6.601x10$^{0}$ | 2.880x10$^{1}$ |
| $\theta_{22}^{(l)}$ | 2 | 1.171x10$^{1}$ | 3.270x10$^{0}$ | 7.961x10$^{-1}$ |
| $\theta_{32}^{(l)}$ | 2 | 1.244x10$^{1}$ | 3.543x10$^{1}$ | 3.355x10$^{1}$ |
| $\theta_{12}$ | 2 | 2.981x10$^{-1}$ | 2.463x10$^{-1}$ | 9.558x10$^{-1}$ |
| $\theta_{22}$ | 2 | 6.164x10$^{-1}$ | 3.631x10$^{-1}$ | 3.521x10$^{-1}$ |
| $\theta_{32}$ | 2 | 3.891x10$^{-1}$ | 5.464x10$^{-1}$ | 1.115x10$^{0}$ |
| $\theta_{2}$ | 2 | 5.000x10$^{0}$ | 5.824x10$^{0}$ | 9.777x10$^{0}$ |
| $\theta_{\epsilon}$ | Noise | 2.693x10$^{-2}$ | 2.693x10$^{-2}$ | 2.693x10$^{-2}$ |
| Log-likelihood | ------ | 3.115x10$^{2}$ | 3.203x10$^{2}$ | 3.141x10$^{2}$ |
| Deviance | ------- | -6.230x10$^{2}$ | -6.406x10$^{2}$ | -6.282x10$^{2}$ |

## C. *Comparison to Benchmark Methods*

The MCGPs for all the scenarios in Fig. 5 are compared to the following four benchmark models:

(i) Independent GP (IGP) for each battery cell with a linear basis function to count aging. The model is trained on the capacity data from the battery cell of interest.

(ii) Average of 100 ANNs with one hidden layer (10 neurons). The cycle number is considered to be the only input and the capacity measurement is considered to be the only output. The model is trained on the capacity data from all the battery cells under the assumption that the cells are identical. Note training the model only on the capacity data from the battery of interest was not effective.

(iii) Average of 100 labeled neural networks (LNN) with one hidden layer (10 neurons). The cycle number and the label are considered to be the inputs and the capacity measurement is considered to be the only output. The model is trained on the capacity data from all the battery cells. The label for the LNNs is encoded with three dummy inputs. Specifically, the dummy input [1,0,0] represents B0005, the dummy input [0,1,0] represents B0006, and the dummy input [0,0,1] represents B0007.

(iv) Average of 100 RNNs with one hidden layer (10 neurons). The cycle number is considered to be the only input and the capacity measurement is considered to be the only output. Unlike the ANN and LNN, the RNN is trained only on the capacity data from the battery cell of interest.

For comparison, the Mean Absolute Error (MAE) and the Mean Squared Error (MSE) for the predictions from the MCGP and the benchmark models are summarized in Table 2. Note that considering more hidden layers for the neural networks will tremendously increase the number of parameters to be estimated. However, that is not desired fore case-studies with limited training data because it leads to poorly trained neural networks that are overfitted and not suitable for predictions.

Table 2. MAE and MSE for the MCGPS and benchmark models for scenarios in Fig. 5.

| Scenario | Fig. 5 (a) | Fig. 5 (b) | Fig. 5 (c) |
|---|---|---|---|
| **MCGP-MAE** | **1.430x10$^{-2}$** | **1.361x10$^{-2}$** | 3.529x10$^{-2}$ |
| **MCGP-MSE** | **2.944x10$^{-4}$** | **2.817x10$^{-4}$** | 1.385x10$^{-3}$ |
| **ANN-MAE** | 2.408x10$^{-2}$ | 1.203x10$^{-1}$ | 1.383x10$^{-1}$ |
| **ANN-MSE** | 7.407x10$^{-4}$ | 1.565x10$^{-2}$ | 1.989x10$^{-2}$ |
| **LNN-MAE** | 3.563x10$^{-2}$ | 2.945x10$^{-2}$ | 4.479x10$^{-2}$ |
| **LNN-MSE** | 1.839x10$^{-3}$ | 1.695x10$^{-3}$ | 2.364x10$^{-3}$ |
| **RNN-MAE** | 2.987x10$^{-2}$ | 5.629x10$^{-2}$ | **2.291x10$^{-2}$** |
| **RNN-MSE** | 1.112x10$^{-3}$ | 4.477x10$^{-3}$ | **7.631x10$^{-4}$** |
| **IGP-MAE** | 1.687x10$^{-2}$ | 1.308x10$^{-1}$ | 2.629x10$^{-2}$ |
| **IGP-MAE** | 4.573x10$^{-4}$ | 1.984x10$^{-2}$ | 1.122x10$^{-3}$ |

Table 2 clearly shows that the MCGP overall outperforms the benchmark models in robustness and accuracy. This is because each of the benchmark methods shows a limitation in at least one aspect. (i) The IGP leverages only the data from the battery cell of interest and that results in poor extrapolation (forecasting) because there is no information about the capacity of future cycles. Therefore, the IGP extrapolation is totally driven by the linear basis function, which failed to forecast the capacity profile of cell B0006. (ii) Individual ANNs showed different predictions and that is a result of the random partition of the training, validation, and testing data. Such instabilities in the performance of individual neural networks indicate that the available data is not sufficient to train a robust ANN. To partially overcome this limitation, the average of 100 ANNs is considered. However, another major limitation for ANNs is ignoring the cell-to-cell variations. This is expected to result in an accurate average performance but not necessarily accurate individualistic performance for each battery cell. The limitation mostly appears for cell B0006. (iii) The average LNN is the most stable among the neural network options because it considers the cell-to-cell variations through the dummy encoded inputs; however, one hidden layer may not be sufficient to capture the nontrivial cross-correlations between the cells. Furthermore, introducing more layers will result in overfitted LNNs due to the increase in the number of parameters to be estimated. (iv) The RNN was expected to outperform the ANN because it feeds back the previously measured/predicted capacity as an input to the network. However, the RNN showed similar performance to the LNN because the RNN is trained only on the capacity data from the battery cell of interest. Finally, the robust and accurate performance of the MCGP is mainly because of the existence of significant nonlinear cross-correlations between the battery cells which were reflected through the latent functions. On average, the MCGP captures the shape of the capacity trend and outperformed all the benchmark methods.

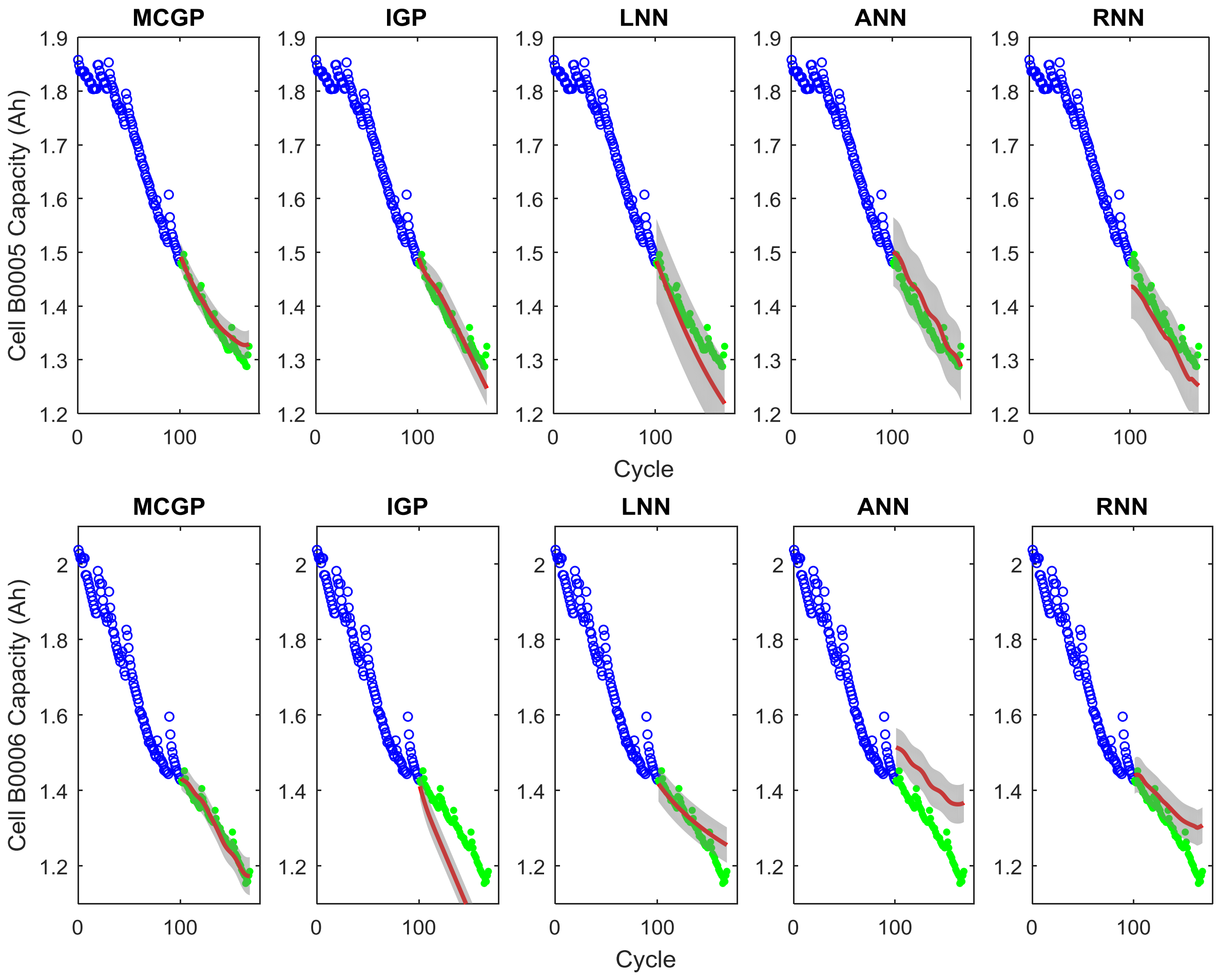

Fig. 6. Long term extrapolation for capacity trends of cells B0005 and B0006 using multiple machine learning models including the proposed MCGP.

Fig. 6 shows the long term forecasting of capacity for battery cells B0005 and B0006. The purpose of the figure is to better visualize the performance of the MCGP in comparison to the benchmark models. The figure strongly validates the results in Table 2 and provides further evidence for the efficacy of the proposed MCGP.

## V. Conclusions & Future Work

The paper proposes a framework that decomposes the capacity profiles of multiple battery cells to latent functions using the MCGP. The latent functions are then convolved with optimized kernel smoothers to reconstruct and predict the capacity of the battery cells. The latent functions leverage the shared cross-correlations between battery cells to better forecast the capacity profile of battery cells that are still at early degradation states. The contributions of the MCGP framework are multifold: (i) it collaboratively models the capacity of multiple battery cells, which enables transferring knowledge between battery cells. (ii) It provides uncertainty information around the predictions, which serves as a robustness-metric for the predictions. (iii) It is non-parametric and does not force a specific function shape on the capacity trends. (iv) It learns nontrivial cross-correlations between the battery cells.

The results show that the proposed MCGP is capable to precisely predict the capacity and capture complex shared trends/patterns between various battery cells as verified experimentally. Furthermore, the proposed method can be used to better estimate the RUL and SOC for Li-ion battery cells, which eventually result in an improved BMS. For future studies, we plan to (i) introduce fixed and dynamic external factors for improved robustness and (ii) integrate the MCGP with newly proposed SOH and RUL models [32]–[36].

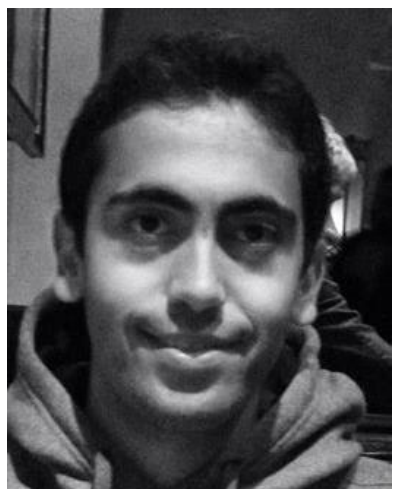

**Abdallah A. Chehade** received the B.S. degree in mechanical engineering from the American University of Beirut, Beirut, Lebanon, in 2011 and the M.S. degree in mechanical engineering, the M.S. degree in industrial engineering, and the Ph.D. in industrial engineering from the University of Wisconsin-Madison in 2014, 2014, and 2017, respectively. Currently, he is an assistant professor in the Department of Industrial and Manufacturing Systems Engineering at the University of Michigan-Dearborn. His research interests are data fusion for process modeling and optimization of data-analysis. Dr. Chehade is a member of INFORMS, IEEE, and IISE.

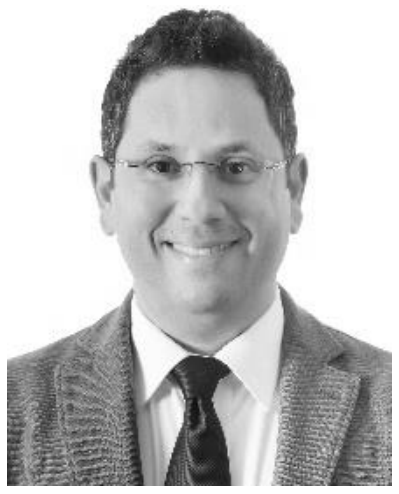

**Ala A. Hussein** received the B.S. degree in electrical engineering in 2005 from Jordan University of Science and Technology, and the M.S. and Ph.D. degrees in electrical engineering from University of Central Florida in 2008 and 2011, respectively.

Dr. Hussein is an Energy Storage and Conversion expert with over 7 years of combined industrial, research and teaching experience. He is a Research Associate with the Florida Solar Energy Center and the Electrical and Computer Engineering Department at the University of Central Florida in Orlando, USA. He has been awarded multiple Research Excellence awards, research grants totaling around $0.4 M and published over 40 papers in leading refereed journals and conference proceedings.